\documentclass[review]{elsarticle}
\usepackage{graphicx}
 \usepackage[usenames]{color}

\usepackage{hyperref}
\hypersetup{colorlinks,allcolors=black}

\usepackage{geometry}
\journal{JINST}
\newcommand{\ada}[0]{ADA}
\newcommand{\adc}[0]{ADC}
\begin{document}

\begin{frontmatter}
%\linenumbers
\title{Performance of ALICE AD modules in the CERN PS test beam}


\author[prague]{M.~Broz}
\author[uas]{J.C.~Cabanillas Noris}
\author[uas]{C.~Duarte Galvan}
\author[catolica]{E.~Endress}
\author[uas]{L.G.~Espinoza Beltr\'an}
\author[fcfm_buap]{A.~Fern\'{a}ndez T\'{e}llez}
\author[russia]{D.~Finogeev}
\author[catolica]{A.M.~Gago}
\author[cinvestav]{G.~Herrera Corral}
\author[korea]{T.~Kim}
\author[russia]{A.~Kurepin}
\author[russia]{A.B.~Kurepin}
\author[russia]{N.~Kurepin}
\author[uas]{I.~Le\'on Monz\'on}
\author[fcfm_buap]{M.I.~Mart\'inez Hernandez}
\author[poland]{C.~Mayer}
\author[finland2]{M.M.~Mieskolainen}
\author[finland1,finland2]{R.Orava}
\author[fcfm_buap]{L.~A.~Perez Moreno}
\author[fermi]{J.-P.~Revol}
\author[fcfm_buap]{M.~Rodr\'iguez-Cahuantzi}
\author[prague,cinvestav]{S.~Rojas Torres}\ead{solangel.rojas.torres@cern.ch}
\author[russia]{D.~Serebryakov}
\author[russia]{A.~Shabanov}
\author[russia]{E.~Usenko}
\author[fcfm_buap]{A.~Villatoro Tello}

\address[prague]{Czech Technical University in Prague, Prague, Czech Republic}
\address[uas]{Universidad Aut\'onoma de Sinaloa, Mexico}
\address[catolica]{Pontificia Universidad Cat\'olica del Per\'u, Lima, Per\'u}
\address[fcfm_buap]{Benem\'erita Universidad Aut\'onoma de Puebla, Mexico}
\address[russia]{Russian Academy of Sciences, Inst. for Nuclear Research, Moscow, Russia}
\address[cinvestav]{Centro de Investigaci\'on y Estudios Avanzados del IPN, Mexico}
\address[korea]{Yonsei University, Seoul, South Korea}
\address[poland]{The Henryk Niewodniczanski Inst. of Nucl. Physics Polish Academy of Sciences, Cracow, Poland}
\address[finland2]{The University of Helsinki, Helsinki, Finland}
\address[finland1]{Helsinki Inst. of Physics, Helsinki, Finland}
\address[fermi]{Centro Studi e Ricerche ``Enrico Fermi'', Roma, Italy}




\begin{abstract}

Two modules of the AD detector have been studied with the test beam at the T10 facility at CERN. The AD detector is made of scintillator pads read out by wave-length shifters (WLS) coupled to clean fibres that carry the produced light to photo-multiplier tubes (PMTs). In ALICE the AD is used to trigger and study the physics of diffractive and ultra-peripheral collisions as well as for a variety of technical tasks like beam-gas background monitoring or as a luminometer.

\par The position dependence of the modules' efficiency has been measured and the effect of hits on the WLS or PMTs has been evaluated. The charge deposited by pions and protons has been measured at different momenta of the test beam. The time resolution is determined as a function of the deposited charge. 
These results are important ingredients to better understand the AD detector, to benchmark  the corresponding simulations, and very importantly they served as a baseline for a similar device, the Forward Diffractive Detector (FDD), being currently built and that will be in operation in ALICE during the LHC Runs 3 and 4.

\end{abstract}

\begin{keyword}
\texttt{Scintillators,} \texttt{Trigger detectors,} \texttt{Performance of High Energy Physics Detectors} 
\end{keyword}

\end{frontmatter}

%\linenumbers
%%%%%%%%%%%%%%%%%%%%%%%%%%%%%%%%%
\section{Introduction
\label{sec:Introduction}}
%%%%%%%%%%%%%%%%%%%%%%%%%%%%%%%%%
ALICE (A Large Ion Collider Experiment)~\cite{Aamodt:2008zz} is one of the four main detectors at the CERN Large 
Hadron Collider (LHC). It is designed to study strongly interacting matter at the highest energy densities reached 
so far in the laboratory, using proton--proton, proton--nucleus and 
nucleus--nucleus collisions~\cite{Carminati:2004fp}. In addition to its main physics program, ALICE is also an 
excellent detector to study other aspects of quantum chromodynamics (QCD) such as diffraction~\cite{Abelev:2012sea}
and photon-induced interactions~\cite{Contreras:2015dqa}.

ALICE started operation in 2009 and has been taken data since then during the so-called Run~1 (2009-2013) 
and Run~2 (2015-2018). Even though during these years the performance of the detector has been 
excellent~\cite{Abelev:2014ffa}, a large part of the current ALICE-Detector setup would not be 
able to cope with the conditions expected at the LHC in Run 3 and 4. 
In order to exploit the increased luminosity and interaction rate in this period, ALICE is now 
implementing a significant upgrade of its detectors and systems~\cite{Abelevetal:2014cna}.

Among the detectors being upgraded is the ALICE Diffractive (AD) detector~\cite{VillatoroTello:2017ccr},
whose new implementation is known as the Forward Diffraction Detector (FDD). Both detectors are composed of 
two arrays installed at each side of the nominal interaction point in ALICE. Each array is made of 4 
sectors, which are made of two layers of identical modules. Each module is made of a plastic scintillator 
pad, wave-length shifters (WLS), optical fibres and photo-multiplier tubes (PMTs). %
Both detectors have the exact same geometry, but the materials of the FDD are faster. %
Here, the main contribution comes from the WLS bar re-emission time that will be reduced from 8.5 ns to 0.9 ns. %
At the same time, the news PMTs have 19 dynodes (instead 16 as the AD PMTs) which will 
reduce after-pulses and also have a more extensive dynamic range. %
Both the AD and the FDD cover the same pseudorapidity ($\eta$) ranges of  $-6.9<\eta<-4.9$ 
and $4.7<\eta<6.3$. Furthermore, the FDD will be also equipped with the possibility of 
having continuous read-out~\cite{Buncic:2011297} in addition to the standard trigger mode. 
The main tasks carried out by the 
AD and to be taken over by the FDD are to participate at the level zero of the trigger system of ALICE, to 
provide physics information for the analysis of diffractive and ultra-peripheral collisions and to 
contribute technical measurements like beam-background monitoring, act as a luminometer, measure the centrality in collisions of heavy nuclei and others.

This article reports the analysis of several test-beam  measurements which were carried out 
with two AD 
modules to determine their characteristics and performance. The data was collected in 2015 at 
the T10 PS 
beam at CERN to study the efficiency, time resolution and charge measurement  of these modules.
These results are not only needed to understand better the response of the AD detector---used in
many ongoing analyses of ALICE data from Run 2---, but also to learn about the
potential of the new FDD. The rest of this article is organised as follows: 
in  Sec.~\ref{sec:ExperimentalSetup} the trigger configuration and the placement of the AD 
detectors in the experimental set-up are described.
Sec.~\ref{sec:Results} reports the results of the analysis of the efficiency, time response 
and charge measurement obtained with the different components of the detector, and when relevant 
as a function of  the  beam energy and and the type of particles---pions and protons--- 
delivered by the beam. Finally in Sec.~\ref{sec:Conclusions}, the conclusions of these studies are summarised.

%%%%%%%%%%%%%%%%%%%%%%%%%%%%%%%%%
\section{Experimental set-up
\label{sec:ExperimentalSetup}}
%%%%%%%%%%%%%%%%%%%%%%%%%%%%%%%%%

\begin{figure}[!t]
	\begin{center}
	    \includegraphics[width=0.7\textwidth]{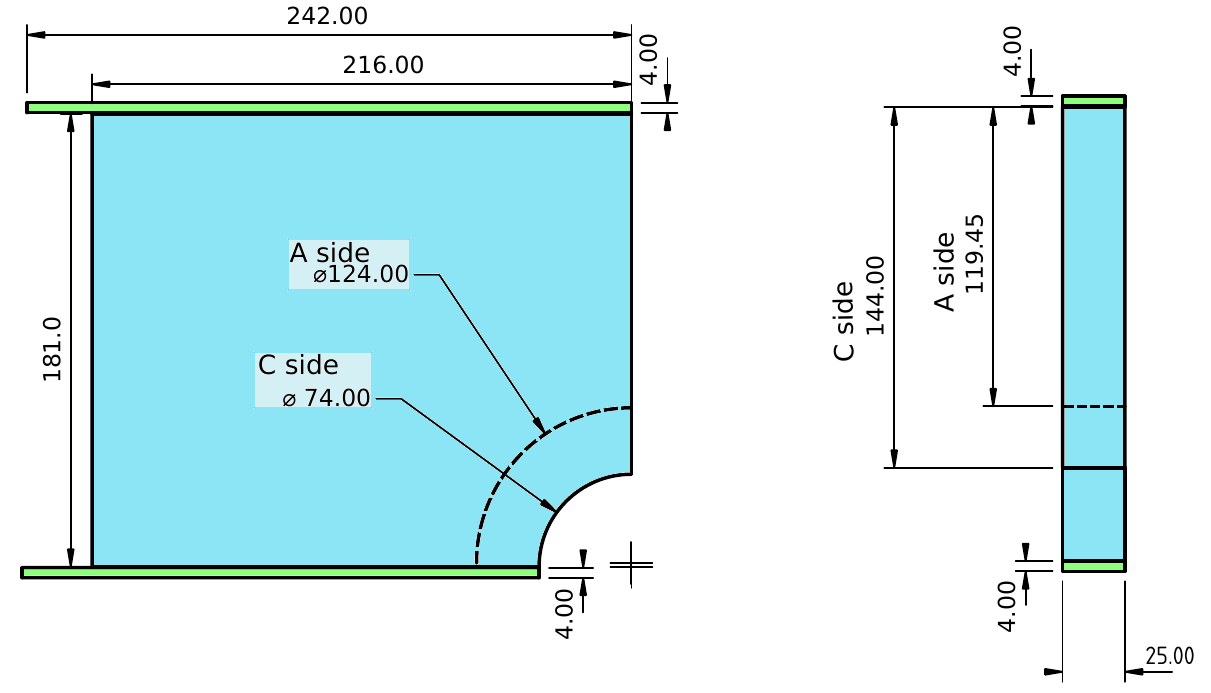}	
		\caption{
		Drawing of the plastic scintillator pads (blue) and  WLS bars (green) of the \ada\, and \adc\, detectors. 
		The difference between the modules is the radius of the cut in the corner to accommodate 
		the beam-pipe: \ada\,(ADC) has a 124 (74) mm radius cut. Therefore the bars attached to 
		the side of the cut, have different lengths: 182 and 207 mm for the \ada\, and \adc\, 
		modules, respectively.
		\label{fig:ADmodules}}
	\end{center}
\end{figure}

Two modules, denoted as \ada\, and \adc, were tested. They  are shown in Fig.~\ref{fig:ADmodules}. The \ada\, and \adc\, modules are identical to the modules used 
in ALICE at positive and negative pseudorapidities, respectively.  These modules are made of 
Bicron BC-404 plastic scintillator, with WLS bars Eljen EJ-208 in two of the sides. % 
The  bars are coupled to clear fibres leading the light to Hamamatsu R5946 PMTs. 
The lengths of the optical fibres were shorter in the test beam, where 47 cm-long fibers were 
used for both modules instead of 100 cm and 54 cm for \ada-type modules, and 250 cm for \adc-type
modules. The different lengths are due to the available space and the placement requirements 
in the ALICE cavern and LHC tunnel, where the detectors are placed, but which were not present 
during the studies with the test beam. %
During the test the PMTs were operated at 1500 and 1650 V for \ada\, and \adc, respectively. 
The read-out was also identical to the one installed in ALICE~\cite{Zoccarato:2011yua}; in 
particular, the signal from the AD modules was split into two signals, one to measure the 
deposited charge and the other, amplified by ten, to determine the arrival time of the particles.

The T10 beam at the Proton  Synchrotron (PS) machine~\cite{Simon:1665434} was used as a source.
The beam consisted mainly of pions ($\pi^{+}$) and protons ($p^{+}$). The following beam momenta
were chosen: 1.0, 1.5, 2.0 and 6 GeV/$c$. 
The relative momentum resolution was 1.3\%.

\begin{figure}[!t]
	\begin{center}
		\includegraphics[width=\textwidth]{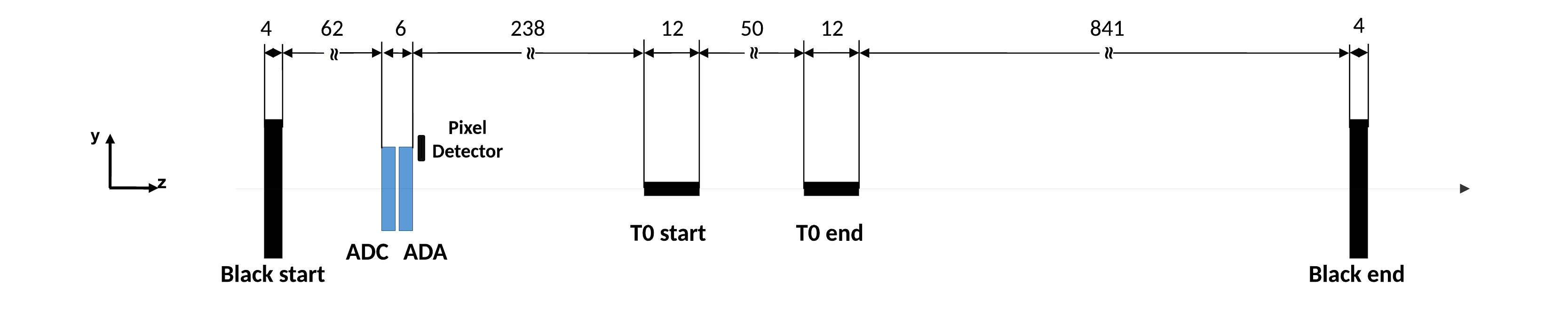}%%
		\caption{
			Test beam set-up in the T10 beam line. The beam direction goes from 
			left to right. All distances are given in centimetres. The drawing is not to scale.
		}
		\label{figure:BeamSetup}
	\end{center}
\end{figure}

The set-up for the data taking in the test beam is shown in Fig.~\ref{figure:BeamSetup}. In addition to the AD
modules, one can also see two other scintillator modules labelled black-start and black-end, as well as two Cherenkov 
radiators denoted as T0-start and T0-end~\cite{Bondila:2005xy}. These devices were used to provide triggers and 
reference timing for the measurements presented below. The specific configurations for these tasks depended on  the 
measurement being carried out and are described in the corresponding sections.

\begin{figure}[!t]
	\begin{center}
		\includegraphics[width=\textwidth]{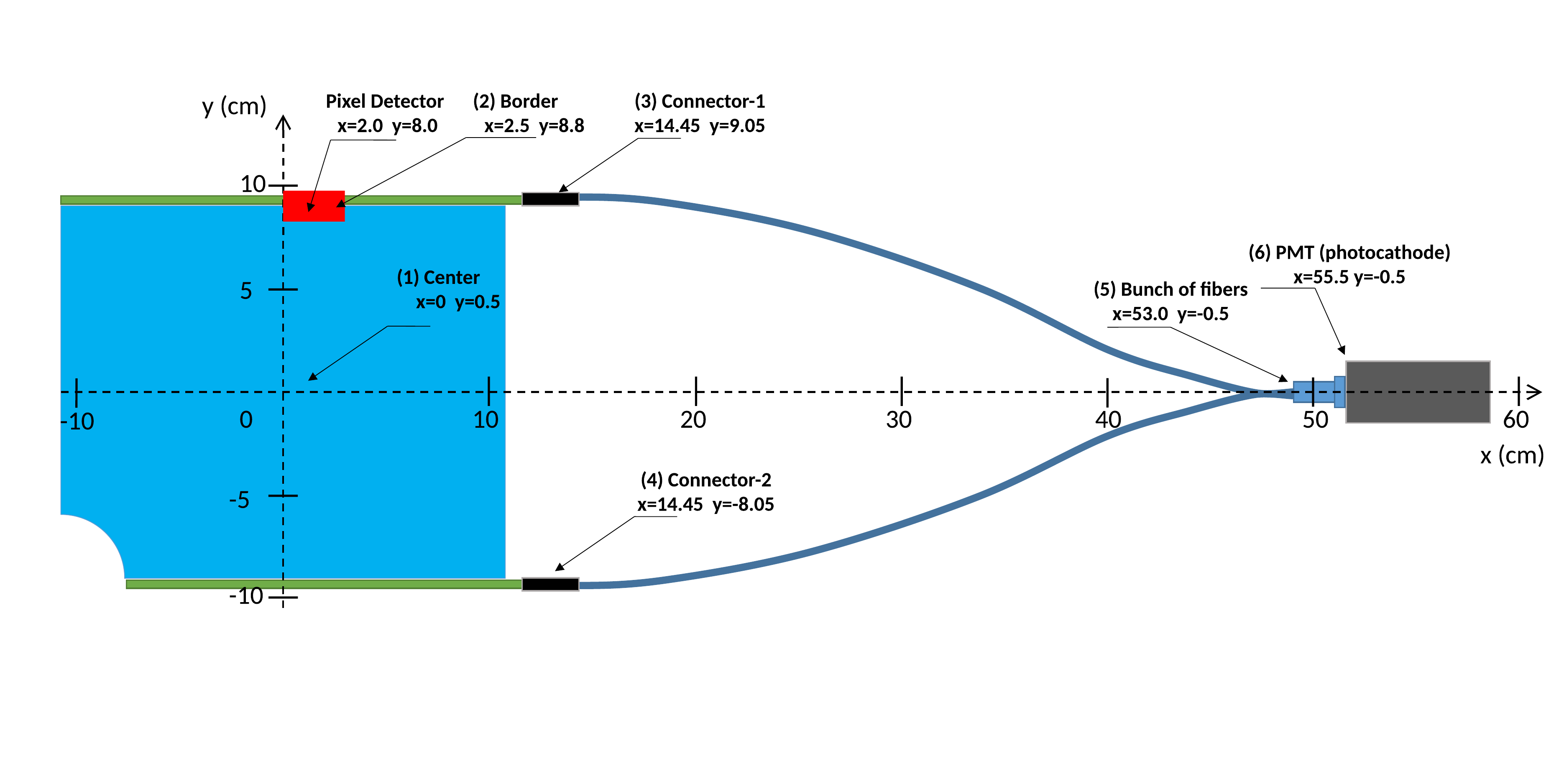}
		\caption{
			Location of the scintillator, the WLS bars, the fibres, 
			the PMT, and the pixel detector, as well as the definition of the coordinate system 
			whose origin lies at the centre of the plastic scintillator.
		}
		\label{figure:zones}
	\end{center}
\end{figure}

The final device involved in the measurements was a silicon pixel sensor positioned at the edge of the scintillator 
pad, as shown in Fig.~\ref{figure:zones}. This detector has an ALPIDE chip as those used for the upgrade of the Inner Tracking 
System~\cite{Abelevetal:2014dna} and for the new Muon Forward Tracker~\cite{CERN-LHCC-2015-001} of ALICE. The chip is based 
on the CMOS MAPS technology; it has  an array of $512 \times 1024$ pixels with a size of 
$26.88 \times 29.24  \,\mu \textrm{m}^2$  and  a sensitive area of $1.376 \times 3.0\, \textrm{cm}^2$. 

%%%%%%%%%%%%%%%%%%%%%%%%%%%%%%%%%
\section{Results
\label{sec:Results}}
%%%%%%%%%%%%%%%%%%%%%%%%%%%%%%%%%
%--------------------------------------------------------------------------------------
\subsection{Efficiency
\label{sec:Efficiency}}

The efficiency  is defined as the fraction of events selected according to  Eq.~(\ref{eq:Eff}):
	\begin{equation}\label{eq:Eff}
		P(\textrm{AD}|\textrm{T0-start} \wedge \textrm{T0-end})
		= \frac{N_{\textrm{AD}}}{N_\textrm{T0}}
		=\frac{ N(\textrm{AD} \wedge \textrm{T0-start}\wedge \textrm{T0-end})}
		{N(\textrm{T0-start}\wedge \textrm{T0-end} )}
	\end{equation}
    where $N_{\textrm{AD}}$ is the number of events that fulfilled the 3-fold 
    coincidence condition, defined by the logic \textrm{AND} between T0-start, T0-end and the AD module, while $N_{\textrm{T0}}$ is the total number of events given by the 2-fold coincidence
    between the T0-start and T0-end. %
    All the triggers are given by the time signal, in particular the \ada\, and \adc\, modules 
    are triggered when the time signal is below a $-40$ mV threshold and inside a 15 ns time 
    window~\cite{Zoccarato:2011yua}. (This time-window in the electronics is  designed to tag 
    beam--beam interaction in the ALICE experiment.)
    The statistical uncertainty is computed as 
    
	\begin{equation}
	\delta=\sqrt{P(1-P)/N_{\textrm{T0}}} \quad \textrm{where} \quad  P=N_{\textrm{AD}}/N_{\textrm{T0}}.
	\end{equation}

The first measurement to be presented is the efficiency to detect the incoming particles as a 
function of the impact position, quantified by the centre of the beam spot. The detector was 
placed over a movable table which could be displaced perpendicular to the beam line such that the 
beam spot could be made to interact at specific positions in the module. 
For these studies the  beam momentum was set at 1 GeV/$c$. 

\begin{figure}[t!]
		\begin{center}
			\includegraphics[width=0.48\textwidth]{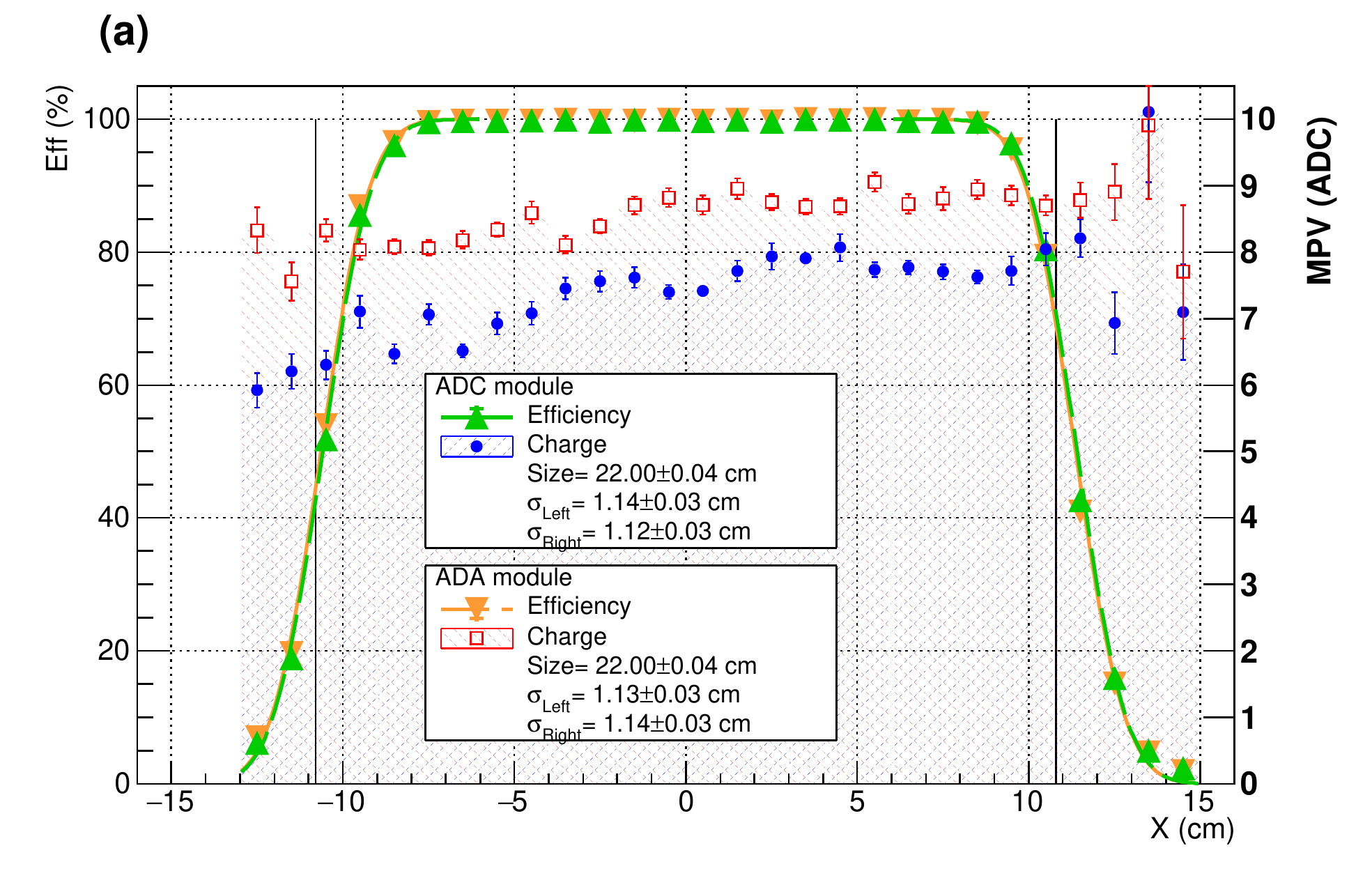}
			\includegraphics[width=0.48\textwidth]{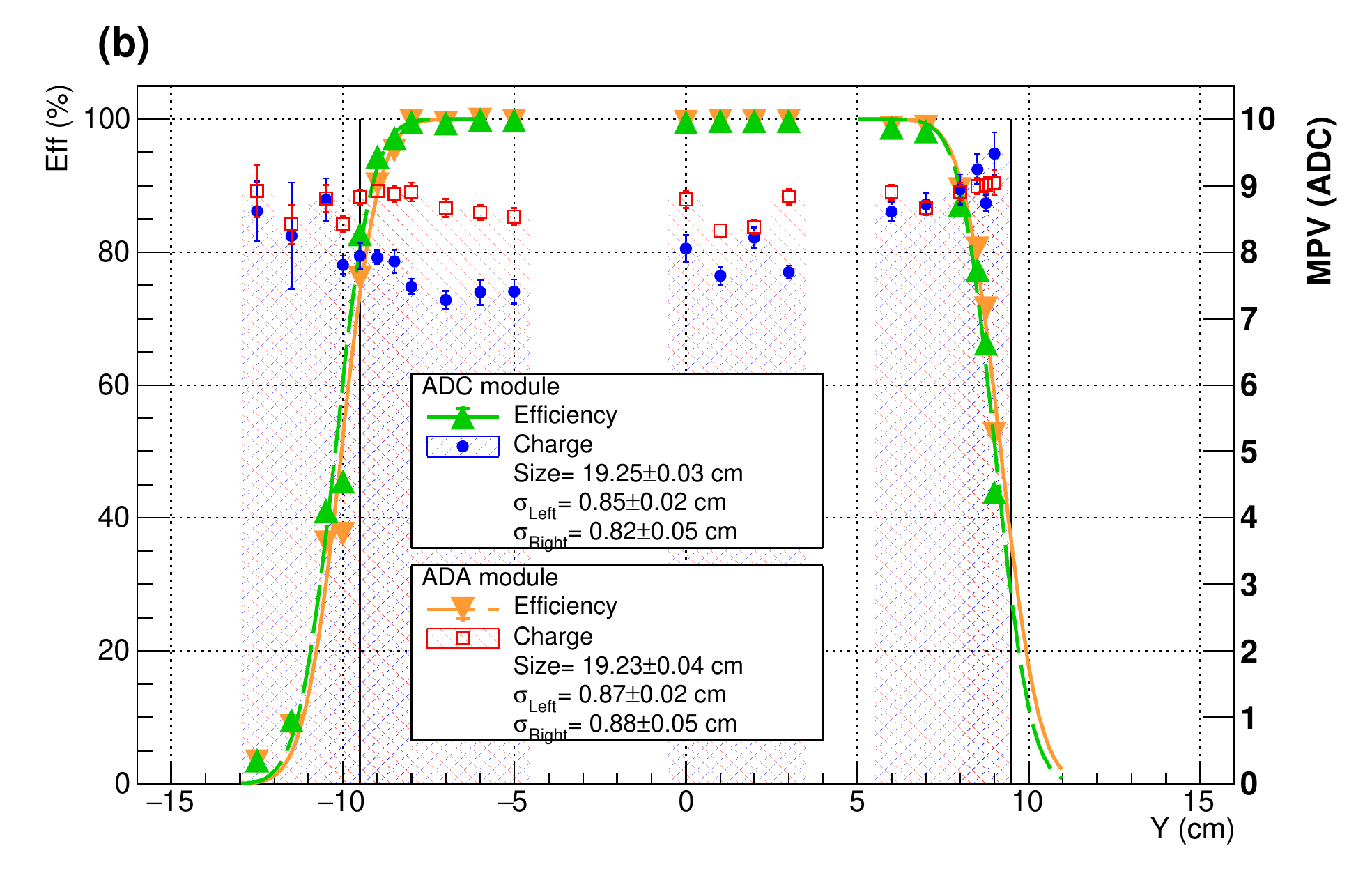}			
			\caption{
			The efficiency, as defined in Eq.~(\ref{eq:Eff}), of the AD 
			modules as a function of the position of the beam spot 
			according to the reference system depicted in 
			Fig.~\ref{figure:zones} is shown with the green and orange 
			symbols and the scale in the left vertical axis. The MPV as 
			defined in Sec.~\ref{sec:ChargeMeasurement} is shown with the 
			red and blue markers and the right vertical axis. See text for 
			details.		}
			\label{figure:Scan-ADcenter}
		\end{center}
	\end{figure}

The results are shown in Fig.~\ref{figure:Scan-ADcenter}. Both detectors are around to $~99.8$\% 
efficient for most of the positions sampled. For the scan on the $Y$ direction some positions were
skipped due to time limitations  to use the beam line and because at that point it was already clear 
that the efficiency is flat also along this axis. The drop of the efficiency at the border and the
existence of an apparent non-zero efficiency outside the nominal acceptance of the module is 
explained by the size of the beam spot. 

To quantify the size of the beam and the corresponding shape of the efficiency curves a	Gaussian Cumulative Function 
distribution (CDF)  is used~\cite{troveCDF}  
	
		\begin{equation} \label{eq:cdf}
	F(X|\mu,\sigma)=\frac{1}{\sigma\sqrt{2\pi}} \displaystyle \int_{-\infty}^{X}
	e^{\frac{-(t-\mu)^{2}}{2\sigma^{2}}} \, dt
	\end{equation}

This CDF is adjusted to the edges of the efficiency curves. The physical length in the vertical and horizontal axes of both 
modules are calculated using the differences between the distances of the mean values of the CDF for each case, obtaining a
length of $X=22 \pm 0.1$ cm and $Y=19\pm 0.1$ cm for each axis, which are consistent with the physical length of the modules  convoluted with the beam size. 
The value of the $\sigma$ parameter for the left and right CDF corresponding to each axis are consistent and are averaged to estimate the beam size of each axis to obtain $\sigma_{Y}=1.13\pm 0.06$ cm and $\sigma_{X}=0.86\pm0.08$ cm.

In order to determine the efficiency at the border of the detector, and to verify that impacts in the WLS bar can be ignored, 
the silicon pixel sensor is used. The  detector is placed so that it covers part of the plastic scintillator and of the WLS bar 
as shown in  Fig.~\ref{figure:zones}. Data taking is triggered by  a coincidence in the black-start and black-end scintillators. 
In this case, the efficiency is defined as: 
	\begin{equation}\label{eq:EffPix}
		P(\textrm{AD}|\textrm{black-start} \wedge \textrm{black-end} \wedge \textrm{Pixel})
		=\frac{ N(\textrm{AD} \wedge \textrm{black-start}\wedge \textrm{black-end} \wedge \textrm{Pixel})}
		{N(\textrm{black-start}\wedge \textrm{black-end} \wedge \textrm{Pixel})}
	\end{equation}
	
\begin{figure}[t!]
	\begin{center}
		\includegraphics[width=\textwidth]{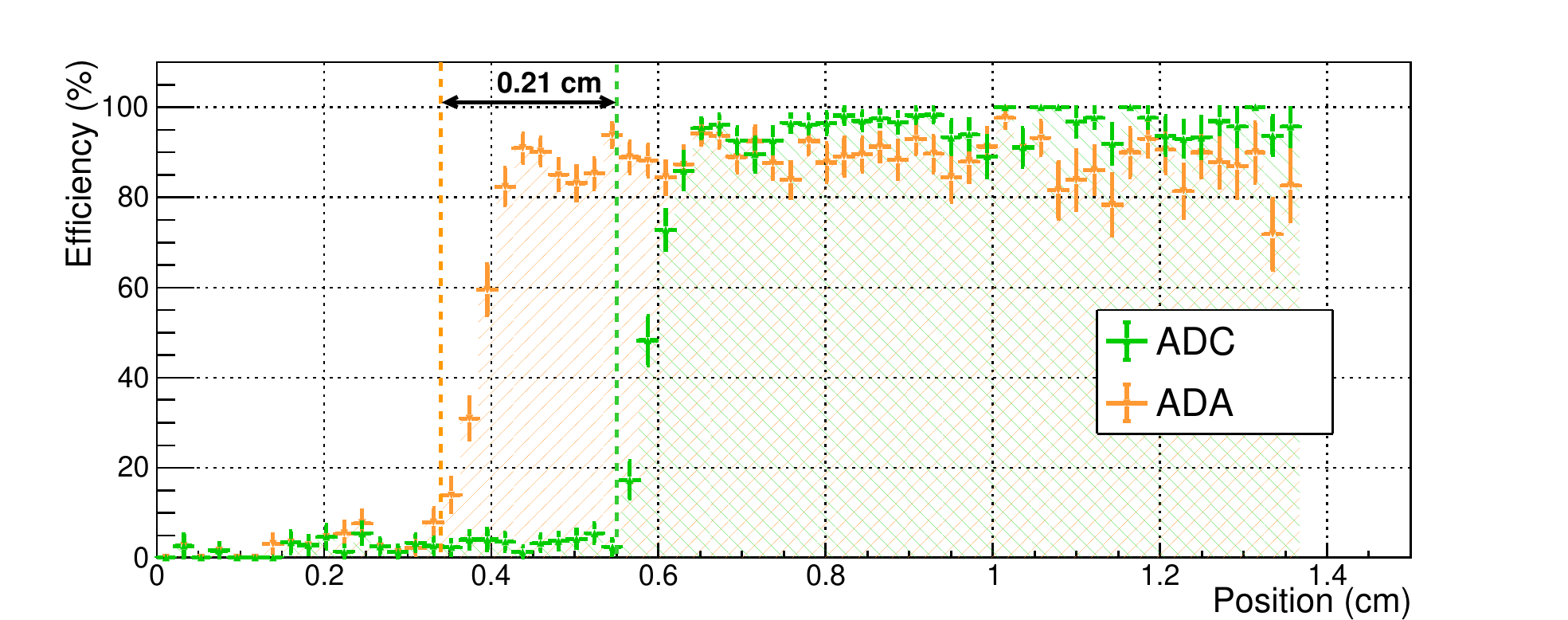}
		\caption{Detection efficiencies along the vertical axis (Y-axis) of \adc\, (green) and 
		\ada\, (orange) near the edge determined with the silicon pixel sensor.}
		\label{figure:effAD}
	\end{center}
\end{figure}

The results are  shown in Fig.~\ref{figure:effAD}. Adjusting a the CDF of  Eq.~(\ref{eq:cdf}) 
the parameters obtained are: $\mu_A= 0.38$ cm, $\sigma_A=0.036$ cm, $\mu_C=0.59$ cm and 
$\sigma_C=0.035$ cm, where the A and C index denote the parameters of the A and C modules.
Additionally,  a misalignment  of approximately 0.21 cm between the two detectors can be seen in
the figure. It is worth to mention that for \adc\, the background shown below 0.55 cm and the
small inefficiency for the places above 0.65 cm are attributed to noisy pixels; the same is valid 
for \ada, but the background is below 0.34 cm and the inefficiencies are above 0.45 cm.

Finally, the efficiency when the beam hits other elements, like the optical connectors, the fibre
bundles or the PMT has also been studied. The efficiency to detect a signal for particles hitting 
the optical connectors (see Fig.~\ref{figure:zones}) is the same in \ada\, and \adc\, and amounts 
to 5\% for the connector at a $Y$ position of $-8.05$ cm, while it raises to 10\% for 
the connector at $Y=9.05$ cm. The efficiency when the fibre bundle at $X=53$ cm is scanned from
$Y=-3$ cm to $Y=3$ cm has a maximum of 15\% (\ada) and 10\% (\adc) at $Y=-0.5$ cm and decays rapidly towards zero within some 1.5 cm. The efficiency when the beam hits at      
$X=55.5$ cm, corresponding to the face of the PMT, is 40\% and decays rapidly with increasing $X$, reaching zero at $X=60$ cm.

%--------------------------------------------------------------------------------------
\subsection{Charge measurement
\label{sec:ChargeMeasurement}}
%--------------------------------------------------------------------------------------
 \begin{figure}[t!]
	\begin{center}
		\includegraphics[width=0.48\textwidth]{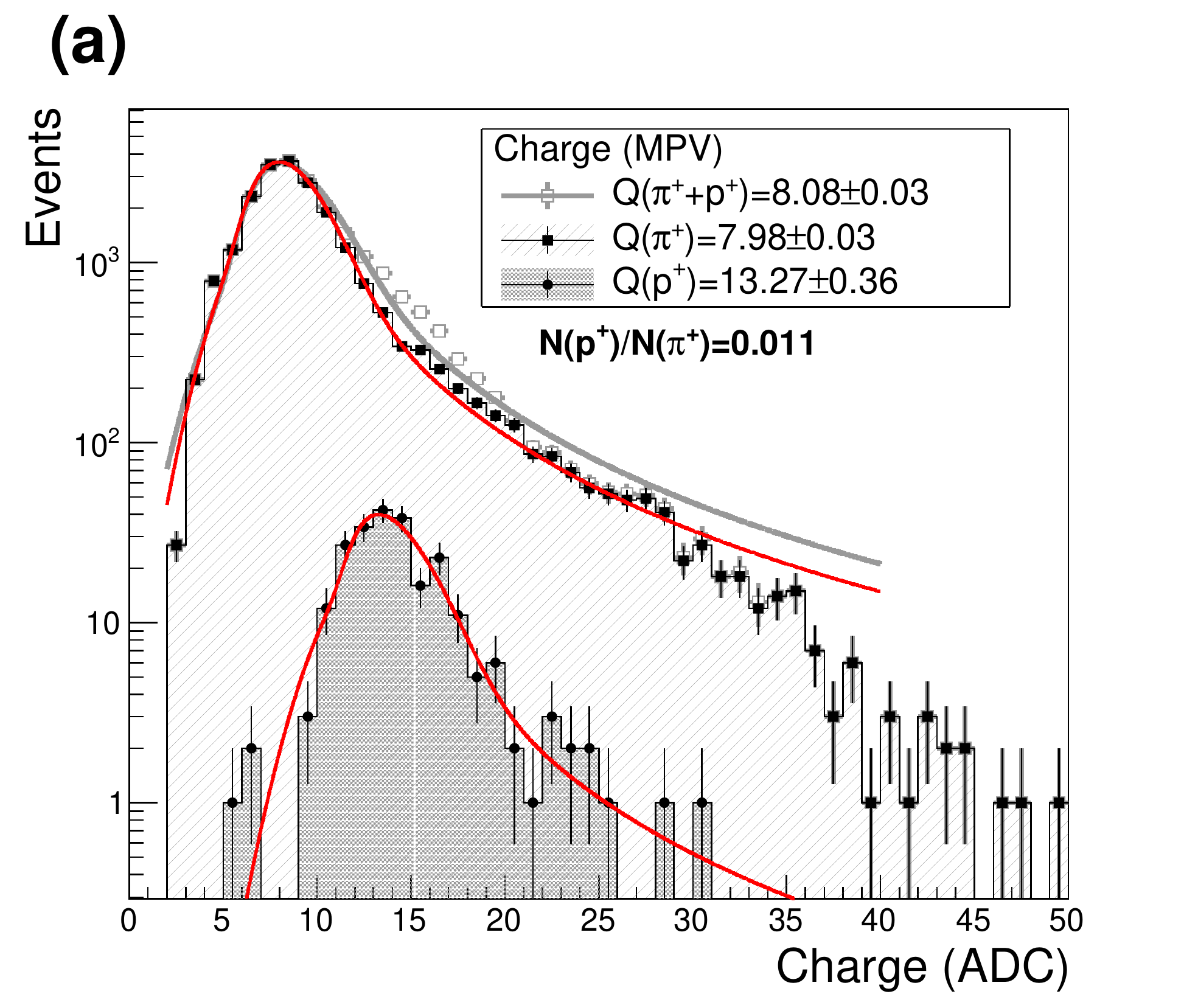}
		\includegraphics[width=0.48\textwidth]{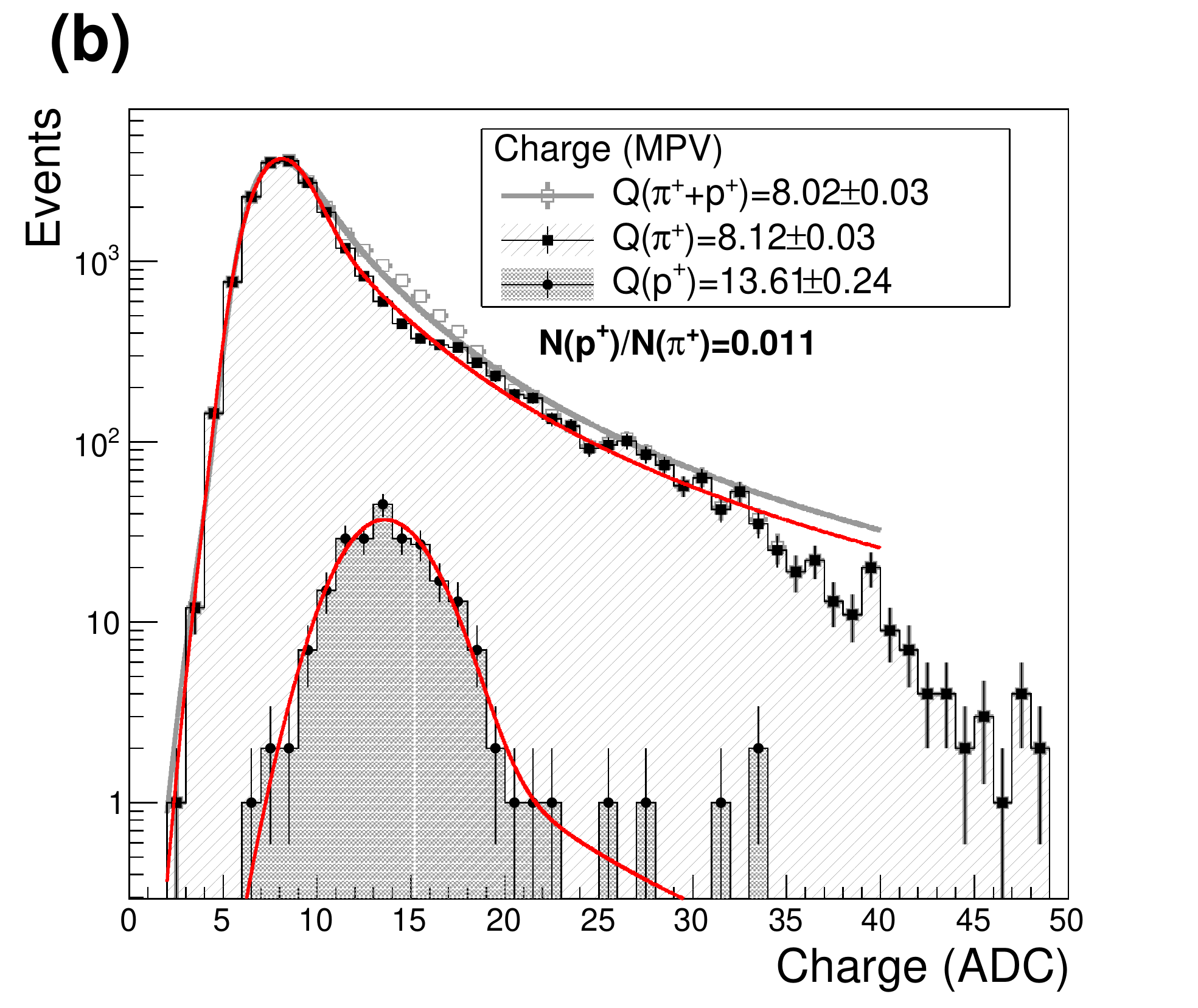}	
		\caption{
			Charge distributions of the (a) \adc\, and (b) \ada\, modules for a 1 GeV/$c$ beam 
			momentum for all particles, and for pions and protons separately. The lines represent 
			a fit to the model mentioned in the text.
		}
		\label{figure:Q_1GeV}
	\end{center}
\end{figure}

 An example of the charge distribution  in the \ada\, and \adc\, modules is shown in 
 Fig.~\ref{figure:Q_1GeV}  along with a model based on the sum of a Landau with a Gaussian 
 distribution, where the most probable value (MPV) of the Landau and mean of the  Gaussian 
 distributions are constrained to be the same. The Gaussian distribution serves to describe  
 detectors effects which smear the charge distribution.
 The data  represented by the light-gray  square markers in Fig.~\ref{figure:Q_1GeV} corresponds to 
 events triggered by the 3-fold coincidence given by $N_{\textrm{AD}}$ in  Eq.~(\ref{eq:Eff}) and 
 were recorded with a beam momentum of 1 GeV/$c$. 
 The most probable value (MPV) is a bit larger  than 8 ADC channels. 

This type of distribution is used to extract the MPV as a function of the impact 
point of the beam spot in the detector. This is shown in Fig.~\ref{figure:Scan-ADcenter} with the 
blue and red markers and the scale in the right-side axes. One observes a slightly different 
value of the MVP in the \ada\, and the \adc\, detectors. This is a geometric effect of the light 
propagation in the scintillator, because the \ada\, detector misses a larger semicircle than 
the \adc\, module as seen in Fig.~\ref{fig:ADmodules}. The slight dependence on the coordinates 
of the impact point of the beam spot is also due to the same  geometric effect.  

The dependences of the MPV on the particle type and on the momentum of the beam have also been 
studied. The first is illustrated in Fig.~\ref{figure:Q_1GeV}, while the latter is shown in 
Table~\ref{table:Qenergies}. The figure shows the case of a beam momentum of 1 GeV/$c$. There is 
a clear separation between the MPV of both particles, where as expected the protons produce 
more charge than the pions. (The particle identification using the time-of-flight technique is 
discussed below.) The table presents the MPV for  both particles as a function of the beam 
momentum. In the case of pions the MPV grows slightly with energy, while for protons the opposite 
behaviour is observed. This behavior is roughly expected, since in this region of the Bethe Bloch curve
the protons go through the  MIP and the pions are already in the relativistic rise.

\begin{table}[t!]
	\centering
	\caption{MPV of the charge measured for pions, protons and the sum of both. For a               momentum of 6 GeV/$c$  it is not possible to 
		    distinguish the particles using the time-of-flight technique described in the text.
	}
	\scalebox{0.87}{
		\begin{tabular}{ | c|c c c|c c c|} \hline
			Momentum& 	\multicolumn{6}{c|}{Charge (ADC counts)} \\ %\hline
			(GeV/$c$) &\multicolumn{3}{c}{\adc} & \multicolumn{3}{c|}{\ada} \\ \hline
			
			&$\pi^+ +p^+$ &$\pi^+$&$\textrm{p}^+$ &$\pi^+ +p^+$&$\pi^+$&$\textrm{p}^+$ \\% \hline
			%% 1 GeV/c resolution
			1.0&8.08$\pm$0.03 &7.98$\pm$0.03 &13.27$\pm$0.036	&8.02$\pm$0.03	&8.12$\pm$0.03 &13.61$\pm$0.24	\\ %\hline
			%% 1.5 GeV/c resolution
			1.5&8.3$\pm$0.04  &8.18$\pm$0.04 &9.72$\pm$0.16 &8.56$\pm$0.05 &8.45$\pm$0.05 &9.94$\pm$0.13	\\% \hline
			%% 2 GeV/c resolution
			2.0&8.21$\pm$0.02 &8.12$\pm$0.02 &8.80$\pm$0.06 &8.41$\pm$0.02 &8.35$\pm$0.02 &8.89$\pm$0.06	\\ %\hline
			6.0& 7.23$\pm$0.09 &- &- &7.14$\pm$0.08 &- &-\\ \hline
		\end{tabular}
	}
	\label{table:Qenergies}
\end{table}

%--------------------------------------------------------------------------------------
\subsection{Time resolution
\label{sec:TimeResolution}}
%--------------------------------------------------------------------------------------

To measure the time resolution one needs two ingredients. First a reference time, and second a 
correction for the the measured  time as a function of the deposited charge, the so-called time 
slewing.
In the set-up shown in Fig.~\ref{figure:BeamSetup} the Cherenkov detectors have the best time 
resolution, below 50 ps~\cite{Bondila:2005xy}, so it is natural to use the T0-end to provide the 
reference time. Comparing the times of T0-end and the black-start detector it is possible to have 
a clean separation of the pion and proton components of the beam as demonstrated below, so that the 
analysis can be done for each particle type separately.
\begin{figure}[!t]
	\begin{center}
		\includegraphics[width=0.4\textwidth]{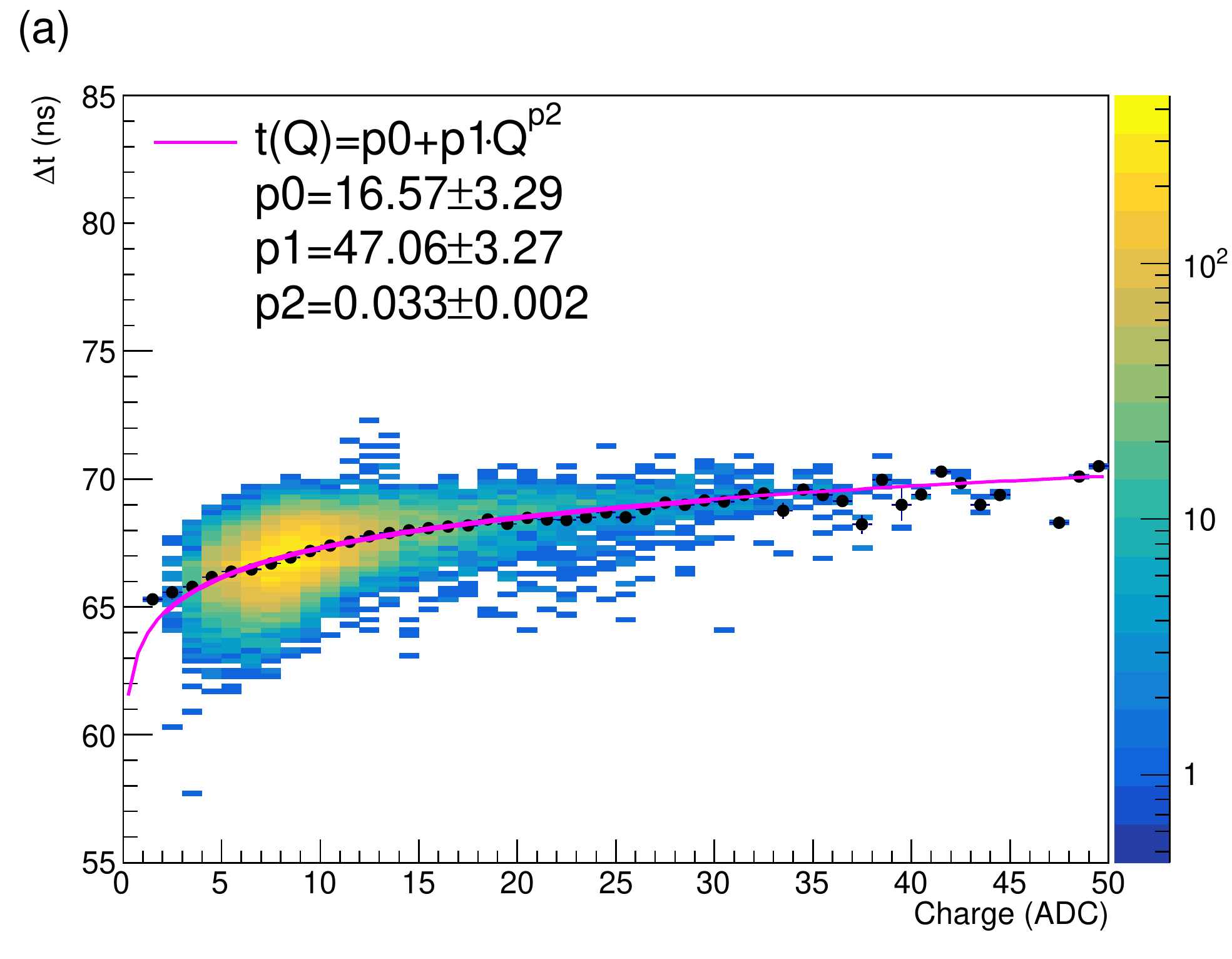}
		\includegraphics[width=0.4\textwidth]{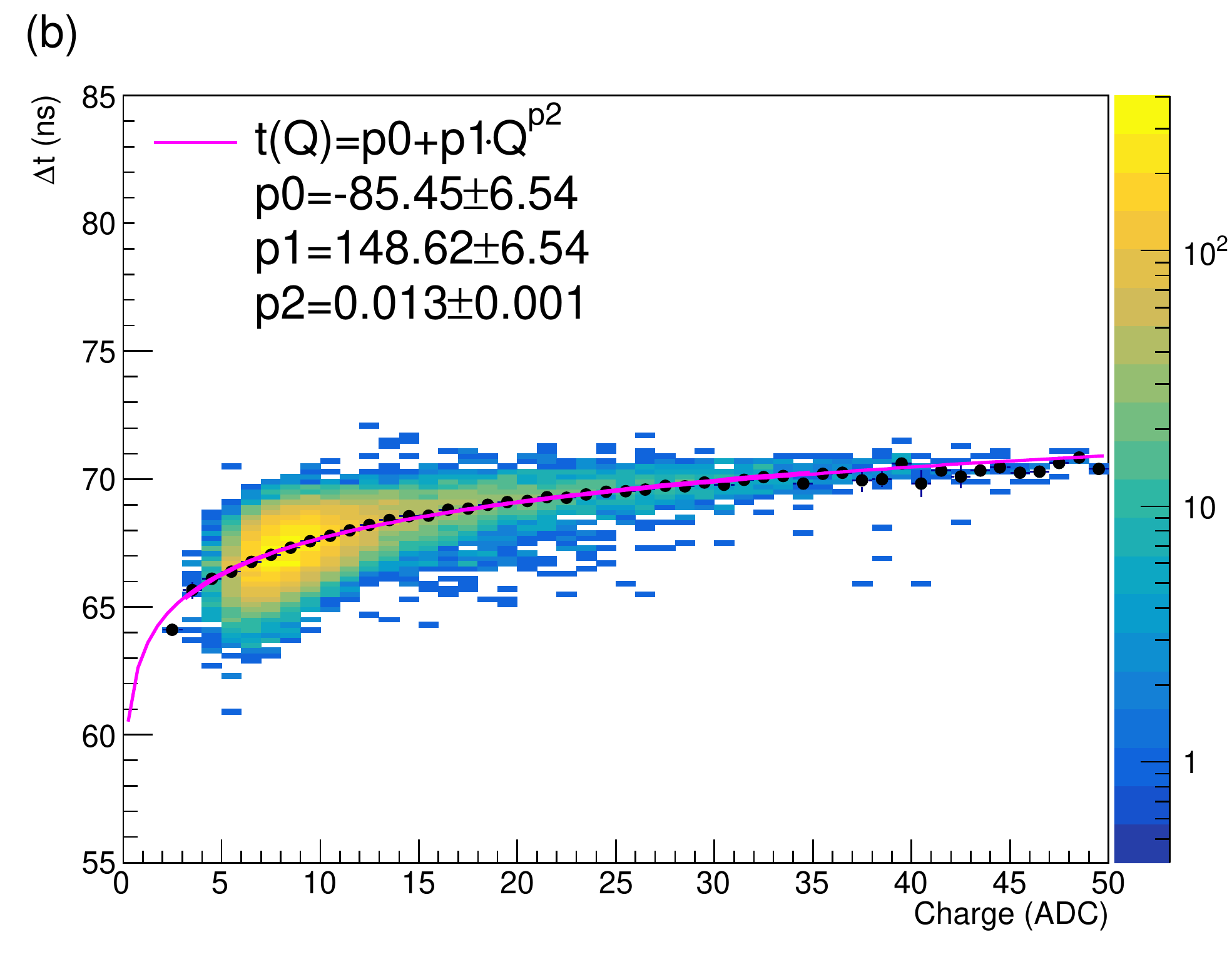}
		\includegraphics[width=0.4\textwidth]{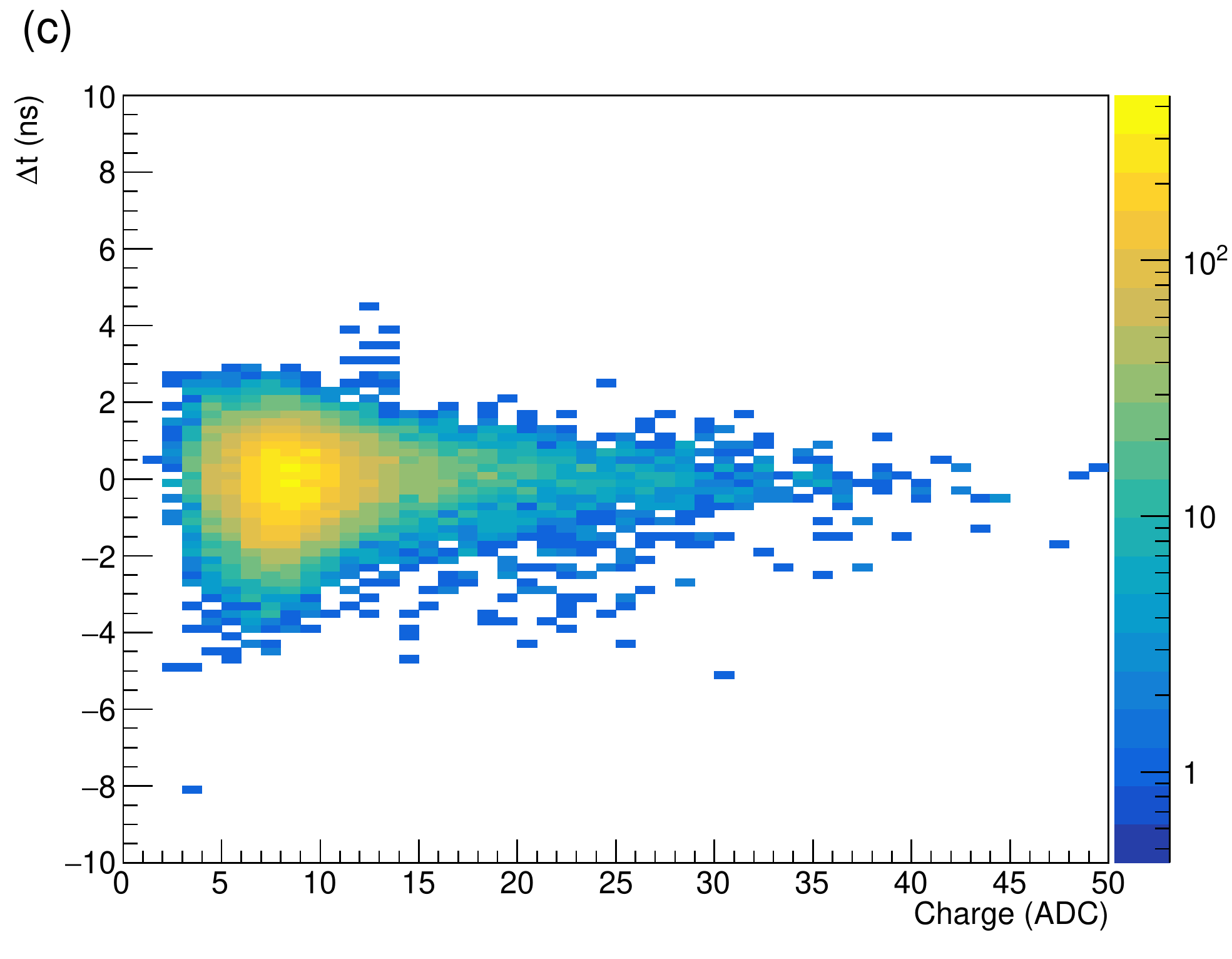}
		\includegraphics[width=0.4\textwidth]{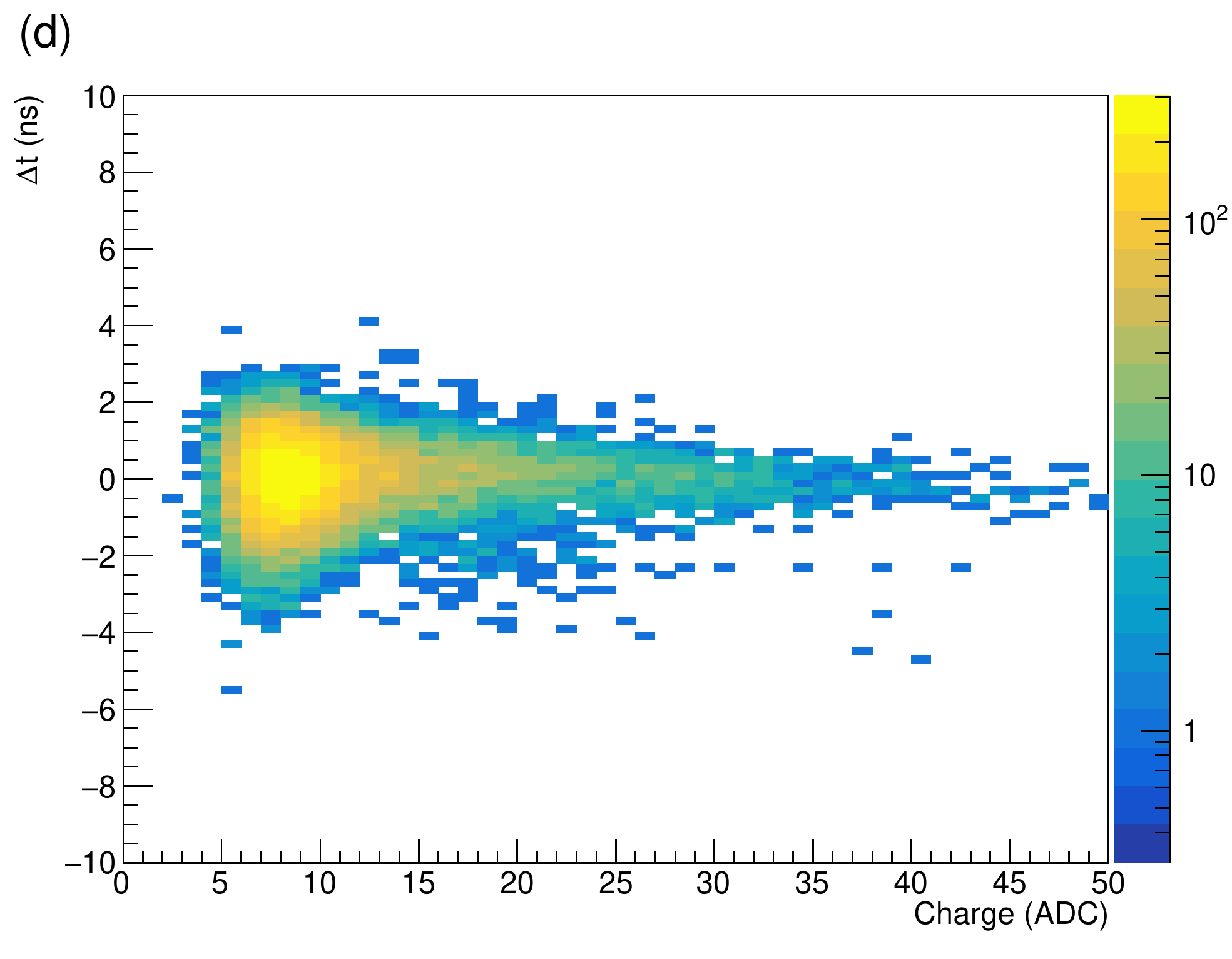}
		\caption{
			Time slewing correction calculated using the pions distributions of \adc\, (a) and \ada\, (b).
			The corrected distributions of \adc\, and \ada\, can be seen in (c) and (d) histograms 
			respectively. 
		}
		\label{figure:TimeQSlew_1GeV}
	\end{center}
\end{figure}

The time measurements are affected by the slewing effect which can be corrected for as follows~\cite{Abbas:2013taa}. The case 
of  pions is used to illustrate  the correction, because the effect is the clearest of both cases.
The (a) and (b) histograms of Fig.~\ref{figure:TimeQSlew_1GeV} show the correlation between the time and the charge measured 
with the \ada\, and \adc\, modules. A clear dependence is seen that can be  parameterised as 
$t(Q)=p0+p1\cdot Q^{p2}$ where $p0$, $p1$, and $p2$ are parameters. The corrected time  is 
calculated subtracting the time $t(Q)$ to the measured time:
$ t(\textrm{corr.})=t(\textrm{measured})-t(Q)$, where $t(\textrm{corr.})$ is the time corrected 
and $t(\textrm{measured})$ is the time measured in each individual event. The corrected distributions
are shown in the (c) and (d) histograms of the same figure. 

\begin{figure}[t!]
	\begin{center}
		\includegraphics[width=0.4\textwidth]{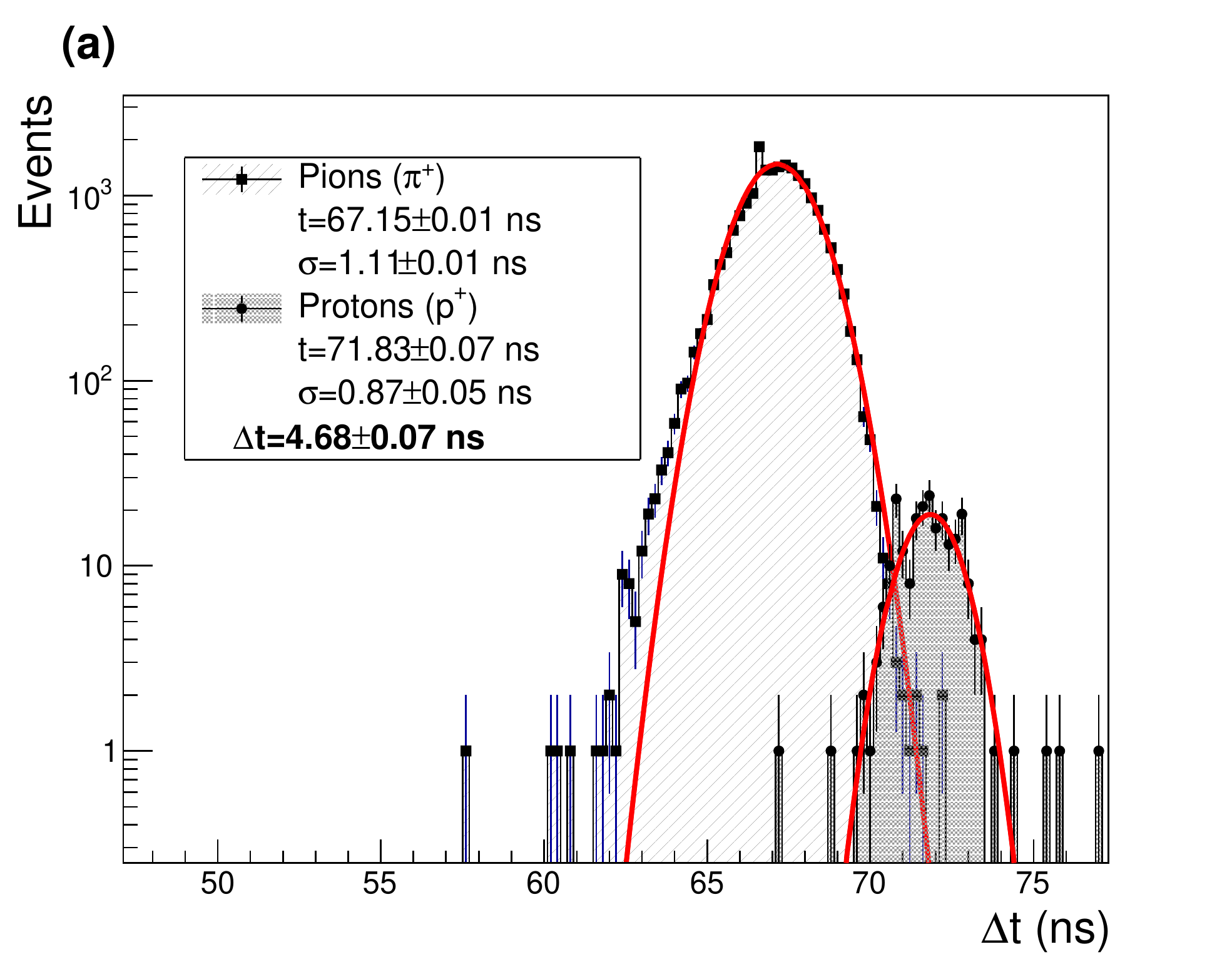}
		\includegraphics[width=0.4\textwidth]{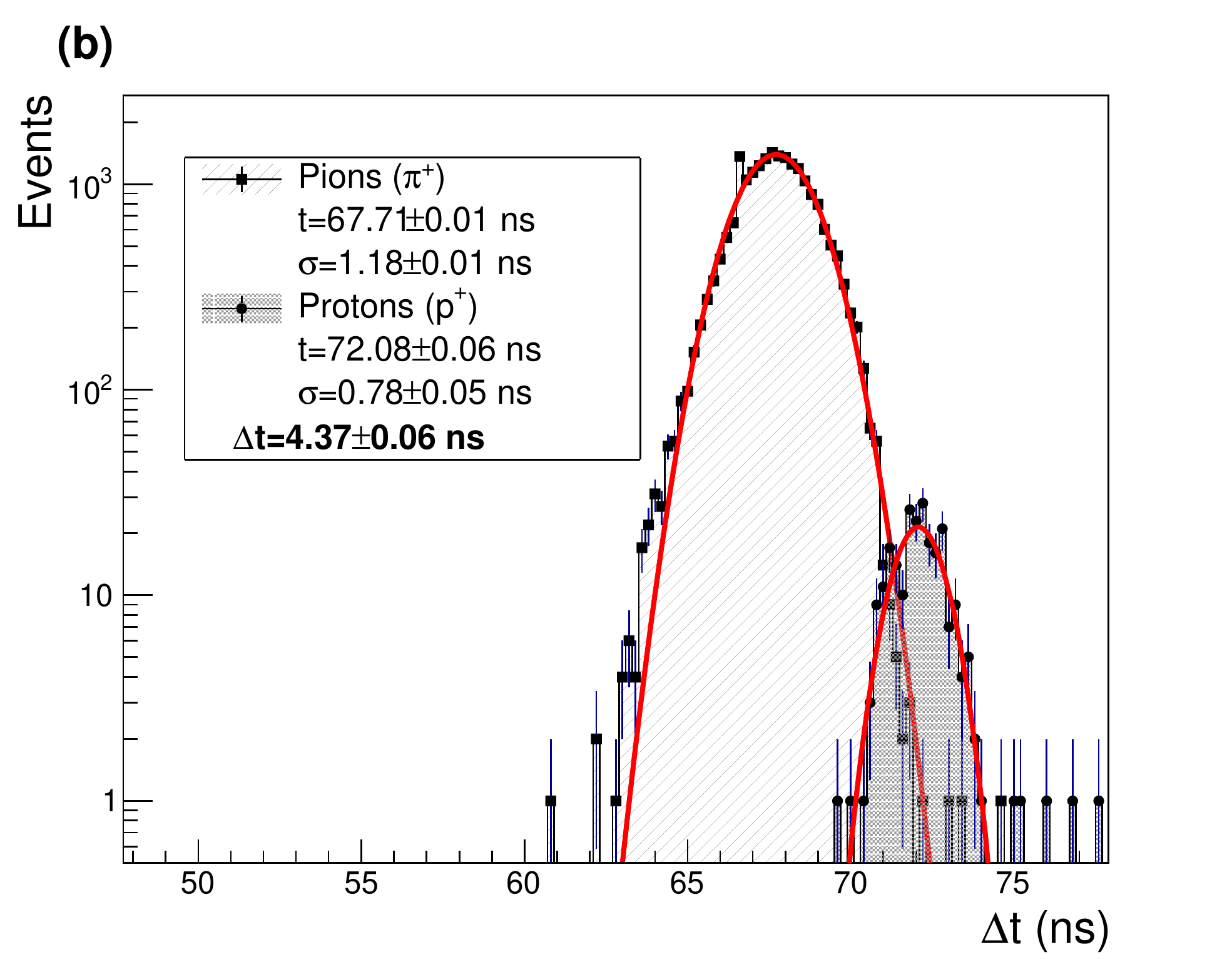}
		\includegraphics[width=0.4\textwidth]{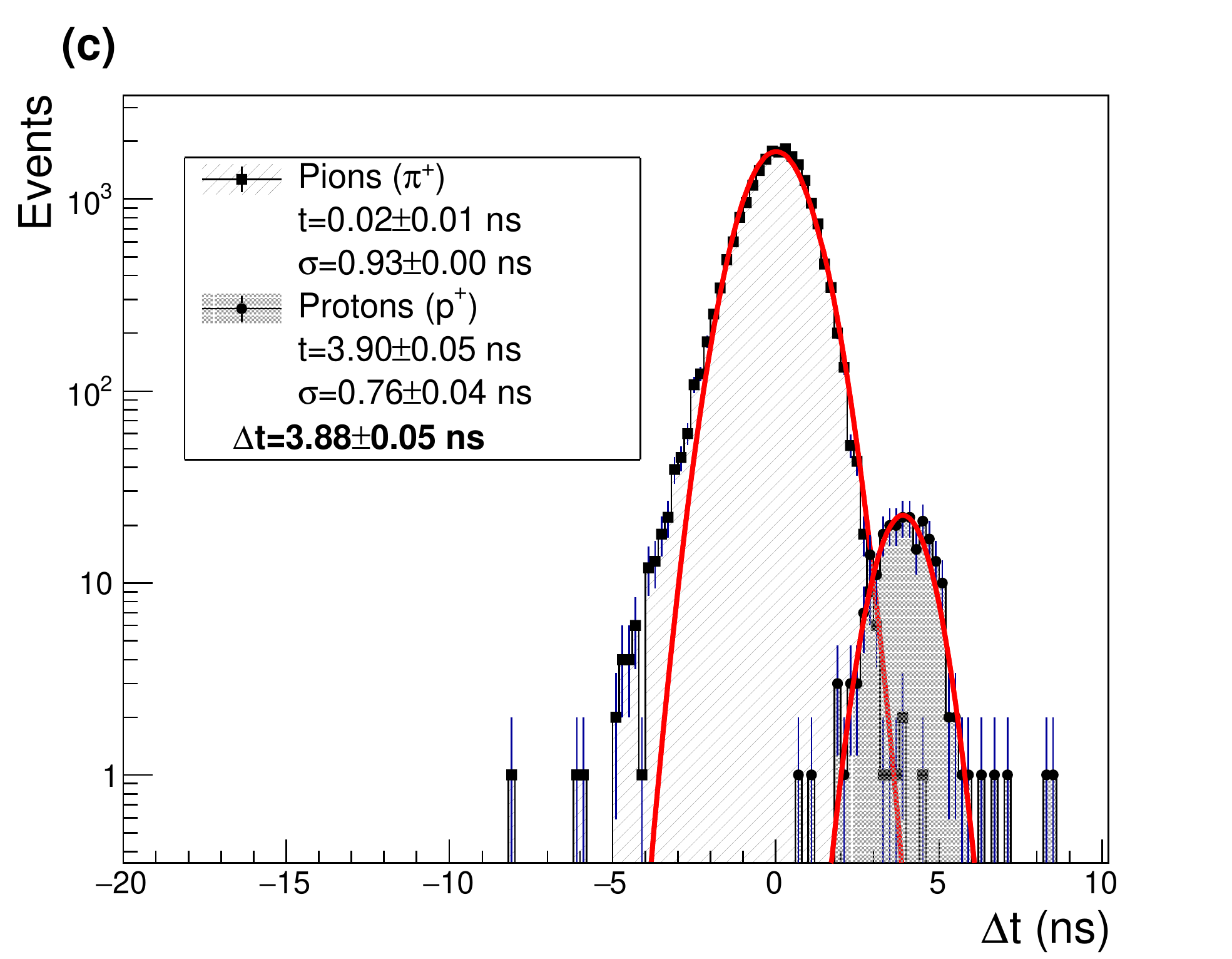}
		\includegraphics[width=0.4\textwidth]{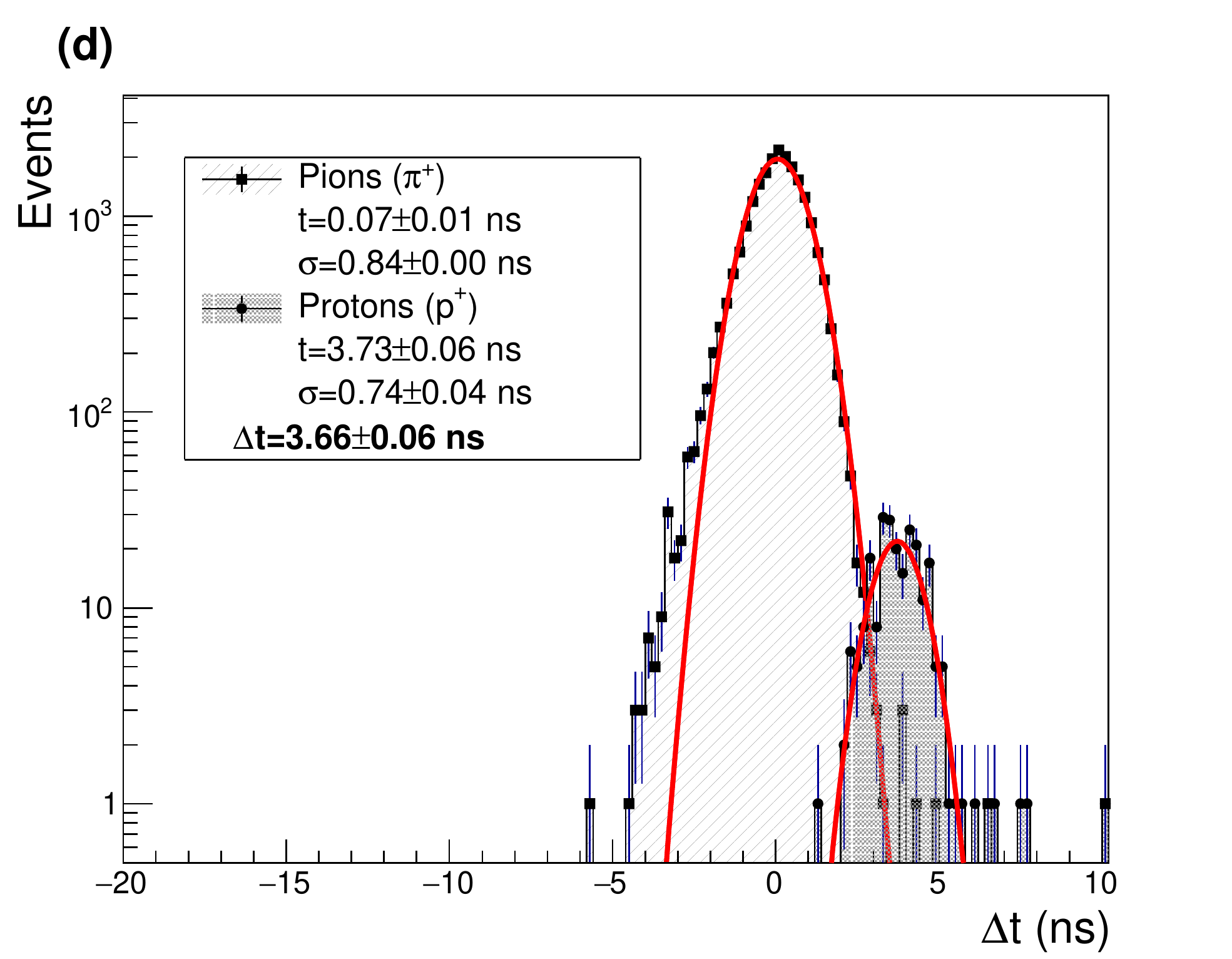}
		\caption{
			Time-of-flight difference  between AD and the T0-end detector 
			($\Delta t= t_{\textrm{AD}}-t_{\textrm{T0-end}}$) for a 1 GeV/$c$ beam momentum,
			and Gaussian fits to the pions and protons contributions.
			The top panels show the uncorrected ADC (a) and ADA (b) time difference. Similarly,
			the bottom row shows the ADC (c) and ADA (d) time differences after applying the 
			time slewing correction. 
		}
		\label{figure:Time_1GeV}
	\end{center}
\end{figure}

The effect that this correction has on the time resolution can be seen in Fig.~\ref{figure:Time_1GeV}
for the case of a beam momentum of 1 GeV/$c$. The mean time difference for both particle species is 
obtained from the difference in mean values of the Gaussian distributions fitted to each 
contribution.

\begin{table}[t!]
	\centering
	\caption{Time resolutions of ADC and ADA at different momenta of pions and protons after the time 
		slewing correction. 
	}
	\scalebox{1.0}{
		\begin{tabular}{ | c|c  c|c  c|} \hline
			Momentum  & \multicolumn{4}{c|}{ $\sigma$ (ns)}  \\ 
			(GeV/c) & \multicolumn{2}{c}{ADC} & \multicolumn{2}{c|}{ADA} \\ \hline
			&$\pi^+$&$\textrm{p}^+$
			&$\pi^+$&$\textrm{p}^+$\\ %\hline
			%% 1 GeV/c resolution
			1.0& 0.93$\pm$0.01	&0.76$\pm$0.04&
			 0.84$\pm$0.01	&0.74$\pm$0.04	\\ %\hline
			%% 1.5 GeV/c resolution
			1.5& 1.26$\pm$0.02	&1.18$\pm$0.07&
			 1.17$\pm$0.02	&1.19$\pm$0.06	\\ %\hline
			%% 2 GeV/c resolution
			2.0 &1.32$\pm$0.01	&1.40$\pm$0.04&
			 1.22$\pm$0.01	&1.33$\pm$0.02	\\% \hline
			6.0&   \multicolumn{2}{c|}{1.12$\pm$0.02} &
			  \multicolumn{2}{c|}{1.18$\pm$0.02} \\ \hline
		\end{tabular}
	}
	\label{table:TimeRes}
\end{table}

A summary of the time resolutions obtained for the different beam momenta is shown in Table~\ref{table:TimeRes}. The resolution 
is similar for both detectors. It increases with momentum and it is larger for pions than for protons. For the lower beam 
momentum, the resolution is below 1 ns and remains well below 1.5 ns even at the largest beam momentum.

\begin{figure}[t!]
	\label{figure:ResCompareExt}
	\begin{center}
		\includegraphics[width=0.75\textwidth]{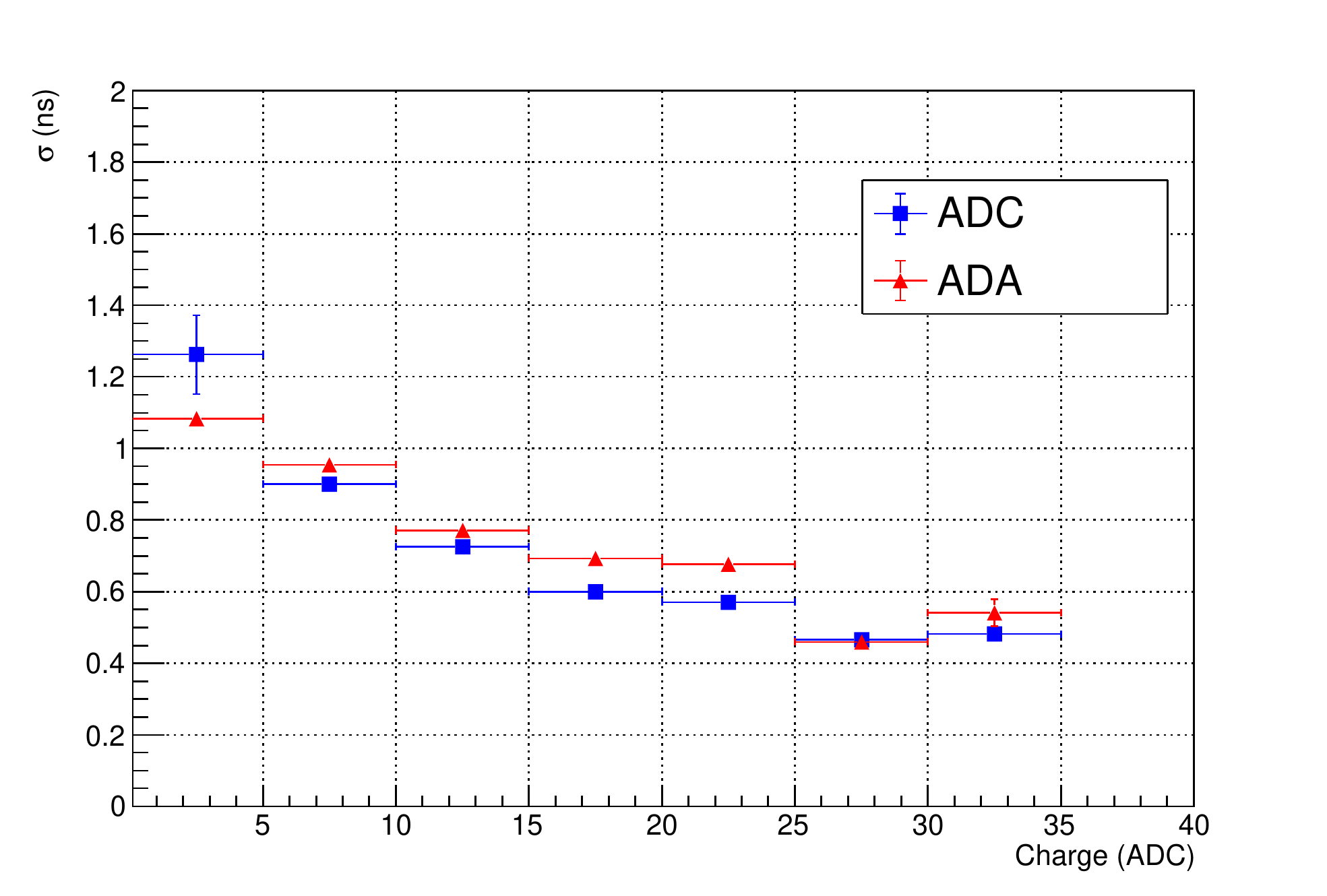}
		\caption{
			Time resolution as a function of the measured charge for the AD modules in the test beam.
		}
	\end{center}
\end{figure}

Finally, the dependence of the time resolution on the deposited charge for pions at 1 GeV/$c$ is 
shown in Fig.~\ref{figure:ResCompareExt}.  
This dependency is obtained dividing the charge spectrum of the $\pi^+$ at 1 GeV/$c$ in seven slices,
each of them corresponds to an interval of 5 ADC counts.
The resolutions correspond to the standard deviations ($\sigma$) of  a Gaussian function fitted 
to the time distribution corresponding to each slice.

%--------------------------------------------------------------------------------------
\subsection{Simple model for the time of flight and the energy deposition
\label{sec:TOF}}
%--------------------------------------------------------------------------------------
Here, a simple model is introduced to describe  the interaction of the particles in the beam with 
the detectors in the experimental set-up. 
First,  the time of flight of a particle is computed  and then compared to the measurements to 
validate the model. Once this is done, the model can be used to obtain the 
energy corresponding to the MPV of the charge distribution.

The first step is to obtain a proper accounting of the time traveled by a given particle while traversing the experimental set-up shown in Fig.~\ref{figure:BeamSetup}. The total distance is
separated in stages: from black-start to \adc\, (65.5 cm), then to \ada\, (3 cm), to T0-start 
(240.5 cm), to T0-end (62 cm), and finally up to black-end (another 854 cm). The distances 
are taken from the centre of the AD and black scintillators pads and the centre of the quartz 
radiator of the T0 detectors, adding 1.5 and 2 cm to the distance with respect to the AD and black 
scintillators respectively; 
for T0 detectors, we use the distance at 1 cm from the edge facing the beam.

At each stage the change of momentum due to the energy loss when traversing the detector materials 
is obtained from a Landau distribution~\cite{Landau:1944if} where the most probable energy loss is:
\begin{equation}\label{eq:landauMPE}
\Delta_p=\xi\left( \ln \frac{2mc^2 \beta^2 \gamma^2}{I}
+ \ln\frac{\xi}{I}+j
-\beta^2-\delta(\beta\gamma) \right)\,\,\,\,    
\textrm{and} \,\,\,\, \xi=\frac{K}{2} \frac{Z}{A}z^2(x/\beta^2).
\end{equation}
	Here  the detector thickness is $x$ in g$\cdot$cm$^{-2}$, $j=0.2$ and $\delta(\beta\gamma)$ is neglected at low energy;
$K=4\pi N_A r_e^2 m_e c^2 = 0.307075$ MeV $\textrm{mol}^{-1}\,\textrm{cm}^2$, $Z$ and  $A$ are the atomic  and mass number
of the material being crossed, $z$ is the charge number of the incident particle, $r_e$ the classical electron radius, $m_e c^2$ the electron 
mass, $N_A$  Avogadro's number, $I$ the mean excitation energy (eV) and $\delta(\beta\gamma)$ the density effect correction 
to the ionization energy loss.

The new energy and momentum of the particle is recalculated after each stage using the material budget  and	$E^2=p^2c^2+M^2c^4$. 
The material budget comprises the following:
for the AD modules 2.5 cm of Bicron 404~\cite{BC404_Manual} each, while the other scintillators are 4 cm thick. The T0 Cherenkov 
radiators have a more
complex composition being made of a 2 cm thick quartz radiator~\cite{Bondila:2005xy} and a PMT. 
The  PMT is modelled by 1 mm of glass (of the vacuum tube), 1 mm of aluminium for the cover at the front and another 1 mm at the 
back; in addition the 16 dynodes are represented by a 0.1 mm thick layer of iron.

Finally, the time after each stage is computed as
\begin{equation}
t_j=\frac{L}{\beta_j c}=\frac{L}{p_j c}\sqrt{p_j^2+m_j^2c^2},
\end{equation}
where $j$ indicates the particle (i.e. pion or proton) and $L$ the distance traveled.
 The time-of-flight difference $\Delta t$ is calculated subtracting the time of flight of the two different particle species traveling the same distance,

\begin{equation}\label{eq:tof}
\Delta t= t_1-t_2= \frac{L}{c}\left[\left(1+\frac{m_1^2c^2}{p^2_1}\right)^{1/2}- 
\left({1+\frac{m_2^2c^2}{p^2_2}}\right)^{1/2} \right].
\end{equation}

Using this model it is possible to compute the expected time differences for the arrival times of a pion and a proton at a 
given detector. The comparison of the model with the measurement is reported in Table~\ref{table:1GeVtimes} for the case of a 
1 GeV/$c$ beam momentum. The agreement is satisfactory for such a simple model of the propagation of the particle beam through 
the experimental set-up.

\begin{table}[t!]
	\centering
	%\resizebox{6cm}{!}{
	\caption{Theoretical and measured time-of-flight differences between pions and protons for the given distances 	with respect to the T0-end detector for a 1 GeV/$c$ beam momentum after the time-slewing correction. 	}
	\begin{tabular}{ | c| c |  c c| } \hline
		Detector & Distance (cm)  &   \multicolumn{2}{c|}{  $\Delta t$ (ns)} \\ 
		&	&Theoretical  &Measurement \\\hline
		
		\adc	    	&305.5	&3.908	&3.9$\pm$0.05	\\
		\ada	    	&302.5	&3.871	&3.66$\pm$0.06	\\
		T0.start 	&62.0	&0.907	&-		\\
		Black.start	&371.0	&4.715	&4.66$\pm$0.03	\\
		Black.end	&845.0	&12.834	&14.5$\pm$0.07\\ \hline
		
	\end{tabular}%}
	\label{table:1GeVtimes}
\end{table}

Once the model has been validated as producing reasonable results for the time-of-flight difference between pions and protons 
it can be used to estimate a conversion factor between the ADC charge and the deposited energy. 
The conversion factor is obtained by dividing the estimated energy deposited in the detector over 
the charge collected for the detector in each measurement (Table~\ref{table:Qenergies}). This was done for pion and proton beams with momenta of 
 1, 1.5 and 2 MeV/$c$. 
The result is shown in Table~\ref{table:AD1_ElossAndWChannel}. The conversion factor is consistent across all cases so it is 
justified to take the average, which yields 

\begin{equation}
\varepsilon=0.832 \pm 0.012 \,\, \textrm{MeV/ADC},
\end{equation}
where the uncertainty is obtained from the standard deviation of the different factors.

\begin{table}[t!]
	\centering
	%\resizebox{6cm}{!} {
	\caption{Theoretical estimation of the energy deposition and energy per ADC count 
		of pions and protons in the ADA and ADC modules, and the corresponding conversion factor of the
	energy deposition estimation with respect to the charge measured.}
	
	%\scalebox{0.86}{
	\begin{tabular}{|c|c c|c c|c c|c c|}
		\hline
		%E_in (MeV)		Eout (MeV)		Momentum (MeV/c)
		Momentum & \multicolumn{4}{ c|}{Theoretical estimation (MeV)}&
		\multicolumn{4}{c|}{Conversion factor (MeV/ADC)}   \\% \cline{2-9}
		(MeV/$c$) & 
		\multicolumn{2}{c}{\adc}& 	\multicolumn{2}{c|}{\ada}& 
		\multicolumn{2}{c}{\adc}&  \multicolumn{2}{c|}{\ada}   \\  \hline%\cline{2-9}
		    &   $\pi^+$ & $p^+$ & $\pi^+$& $p^+$ &   $\pi^+$ & $p^+$ & $\pi^+$& $p^+$ \\ 
		1000&   6.69&	10.89&  6.68&	10.95&  
		        0.838$\pm$0.003 &0.820$\pm$0.002&	0.823$\pm$0.003     &0.805$\pm$0.014\\
		1500&   6.85&	8.14&   6.85&	8.16&  	
		        0.838$\pm$0.004 &0.837$\pm$0.014&   0.8107$\pm$0.005    &0.821$\pm$0.0107\\
		2000&   7.00&	7.25&   6.99&	7.26&	
		        0.862$\pm$0.002 &0.824$\pm$0.006& 	0.838$\pm$0.002     &0.816$\pm$0.006\\
		6000&   7.63&	6.65&   7.63&	6.65&
		        -&  -& 	-&  -\\ 
		\hline
	\end{tabular}%}
	%}
	\label{table:AD1_ElossAndWChannel}
\end{table}

%%%%%%%%%%%%%%%%%%%%%%%%%%%%%%%%%
\section{Summary and outlook \label{sec:Conclusions}}

One \ada\, and one \adc\, modules have been studied with test beams. The set-up allows for the 
measurement of the modules' efficiencies as a function of the beam impact position, and the 
measurement of the charge deposition as well as the time resolution of the modules. In particular, 
these two last properties have been studied as a function of the particle species, pion or proton, 
and the momentum of the test beam. A simple model of the energy deposition on the material of the 
set-up traversed by the beam particles allows for the conversion of the ADC charge into an energy.  

The results show that the modules have a high and uniform efficiency as a function of the impact 
point and that the hits in other parts of the detector, like the WLS or the PMT, have a lower 
efficiency which decays steeply outside of a very small range of positions. The set-up allows to 
separate the pion and proton components of the beam. The charge deposition in both cases has been 
measured and shown to behave as expected. The time resolution has been measured and shown to be 
below 1.5 ns for particles depositing little charge in the modules and below 1 ns for particles 
depositing more charge.

These results allow for a better understanding of the detector and provide key information to 
benchmark and improve the simulation of the AD detector used during the LHC Run 2 in ALICE. This 
knowledge can be applied in a straightforward manner to the the FDD being constructed now and that 
will be in operation in ALICE during the LHC Runs 3 and 4. 

%%%%%%%%%%%%%%%%%%%%%%%%%%%%%%%%%

%%%%%%%%%%%%%%%%%%%%%%%%%%%%%%%%%
\appendix
%%%%%%%%%%%%%%%%%%%%%%%%%%%%%%%%%

%--------------------------------------------------------------------------------------
%\acknowledgments
%--------------------------------------------------------------------------------------

\section*{Acknowledgments}
We thank  Paolo Martinengo from the ALICE Inner Tracking System for allowing us the use of the 
ALPIDE pixel sensor chip for our studies in the T10 PS beam at CERN.
This work was partially supported by Consejo Nacional de Ciencia y Tecnolog\'ia (Mexico) grant number A1-S-13525 and the 
Ministry of Education, Youth and Sports of the Czech Republic under the grant number LTT17018. 


%%%%%%%%%%%%%%%%%%%%%%%%%%%%%%%%%
\bibliography{ref}
%\bibliographystyle{unsrt} 
%%%%%%%%%%%%%%%%%%%%%%%%%%%%%%%%%

\end{document}